\documentclass[twocolumn,showpacs,preprintnumbers,amsmath,amssymb,prl]{revtex4}
\usepackage{graphicx}
\usepackage{fancyhdr}
\usepackage{dcolumn}
\usepackage{bm}
\usepackage{epstopdf}
  \usepackage{natbib}
  \usepackage[a4paper,left=1.55cm,top=2cm,right=1.55cm]{geometry}
\usepackage{amssymb}
\usepackage{amsmath}
\usepackage{graphicx}
\usepackage[normalem]{ulem}
\usepackage[dvips]{color}

\begin{document}
\title{  Water Transport through Carbon Nanotubes with the Radial Breathing Mode}
\author{Qi-Lin Zhang$^{1,2}$, Wei-Zhou Jiang$^{1}$\footnote{wzjiang@seu.edu.cn},
Jian Liu$^{3}$, Ren-De Miao$^{4}$, Nan Sheng$^{3}$ }
\affiliation{  $^1$ Department of Physics, Southeast University, Nanjing 211189, China\\
$^2$ Department of Mathematics and Physics, Anhui Polytechnic University, Anhui 241000, China\\
$^3$Shanghai Institute of Applied Physics, Chinese Academy of Sciences, Shanghai 201800, China\\
$^4$Institute of Science, PLA University of Science and Technology, Nanjing 211101, China}

\begin{abstract}
Molecular dynamics simulations are performed to investigate the water
permeation across the single-walled carbon nanotube (SWCNT) with the
radial breathing mode (RBM) vibration. It is found that the RBM can
play a significant role in breaking hydrogen bonds of the water
chain, accordingly increasing the net flux dramatically, and reducing
drastically the average number of water molecules inside the tube
with the frequency ranging from 5000 to 11000 GHz, while far away
from this frequency region the transport properties of water
molecules are almost unaffected by the RBM. This phenomenon can be
understood as the resonant response of the water molecule chain to
the RBM. Our findings are expected to be helpful for the design of
high-flux nanochannels and the understanding of  biological
activities especially the water channelling.
\end{abstract}
\pacs{31.15.at, 31.15.xv, 61.20.Ja}
\maketitle

\emph{Introduction}---The transport properties of water molecules
across nanochannels have been  extensively investigated in the
past~\cite{mur00,zhu04,gon08,son08,liu10,duan10,hol06}. In recent
years, various designs of nanochannels concerning temperature
gradients~\cite{lon07}, charge modification~\cite{li07}, coulomb
dragging~\cite{wang06}, and static electric fields~\cite{jia11} have
been proposed to study abundant novel properties of water permeation.
These progresses may have significant technical implications for the
design of novel nanofluidic devices or machines, such as desalination
of seawater~\cite{corr08}, molecular sieves~\cite{jira97}, molecular
water pump~\cite{gon07}, and so on. Furthermore, one important reason
why research on nanochannel transport properties has continued to be
of interest is that some primary characters  of water confined in
simple nanochannels are similar to that of complex biological
channels. Typically, water across both the biological membrane
proteins (for instance, aquaporin-1 (AQP1) and Glpf)~\cite{groo01}
and  the nanochannels~\cite{hu01} can form single-file water chains
of  stable dipole orientation. The vapor-liquid transition has also
been observed in mechanosensitive membrane channels~\cite{anis04} and
other hydrophobic nanopores~\cite{beck03}. Another resemblance
existing in the nanochannels and biological channels is that they can
share both wavelike density distribution and wet-dry transition
resulting from confinement~\cite{wan05,lot1}.

It has been long well-known that the wave mechanics well interprets
the truth of nature. Microscopic objects such as biological
microtubules are actually exposed in the environmental vibration
modes of all possible micro-oscillators entangled with intrinsic
modes. It is thus of broad interest to study the molecular
permeations with the vibrational coherence and/or interference.
Indeed, theoretical and experimental studies indicate that the
single-walled carbon nanotubes (SWCNTs) have various collective
vibrational modes~\cite{rao97,dub03,law05,yan07}. Collective motion,
different from the internal fast processes, is usually regarded as
the fifth dimensional motion, whereas it remains almost untouched in
the study of water permeation through SWCNTs. The radial breathing
mode (RBM)~\cite{arau10}, being the simplest collective vibration, is
one of the most important modes of SWCNTs, and it can be excited
easily by radio wave techniques or constantly by environmental
interferences. In this study, we thus restrain ourselves to the RBM.
It is known that molecular dynamics (MD) simulations have been widely
used to investigate the transport properties of water molecules
across SWCNTs~\cite{zhu04,gon08,wan05,hu01,zuo10}. In this work, we
use MD simulations to explore the dynamic behavior of water
permeation through SWCNTs with the RBM. We have observed that the RBM
of the SWCNT can greatly affect the water permeation properties
through a resonant manner.

\begin{figure}[thb]
\begin{center}
\includegraphics[height=4.5cm,width=7.0cm]{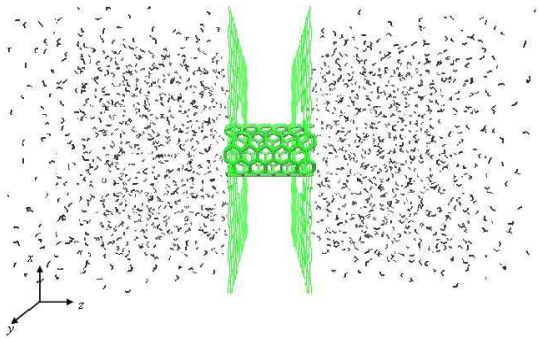}
 \end{center}
\vspace*{-5mm} \caption{(Color online) Snapshot of the simulation
system. An uncapped (6,6) armchair  SWCNT combined with two graphite
sheets is solvated in a water box of
3.5$\times$3.5$\times$6.3nm$^{3}$ with 1852 molecules. There is a
vacuum between two sheets. The axis of SWCNT is parallel to the
z-axis.\label{fel}}
\end{figure}

\emph{Computational Methods}--- The simulation framework is
illustrated in Fig.~\ref{fel}. An uncapped (6,6) armchair SWCNT with
a length of 1.34 nm and a diameter of 0.81 nm is embedded in two
graphite sheets along the z direction. The distance between the
bottom end of the SWCNT and the graphite sheet is 2
$\overset{\circ}{A}$.  Periodic boundary conditions are applied in
all directions. The electrostatic calculation, carried out at a
constant volume with box size dimensions of $L_x=3.5$ nm, $L_y=3.5$ nm,
$L_z=6.3$ nm and constant temperature (300 K) achieved by Langevin
dynamics, is done using the PME~\cite{dar93} summation method with a
cutoff for real space of 1.2 nm. The cutoff for the van der Waals
interaction is also 1.2 nm. With the MD program NAMD2~\cite{jam05},
all simulations are performed using the CHARMM27 force
field~\cite{mac98,zhu02} and  the TIP3P water model~\cite{jor83}. The
time step of 1fs is used, and data are collected every 0.5 ps. To
obtain a directed flow, an additional acceleration of 0.01
nmps$^{-2}$, equivalent to a pressure difference about 20 Mpa between
two ends of the SWCNT, is applied to each water molecule along +z
direction~\cite{zhu02,gon08}.

The RBM is simply simulated with a coherent radial motion obeying the
following relation
\begin{equation}\label{eqsym}
r(x,y,t)=r_0(x_0,y_0,0)+A{\cos}({\omega}t+{\varphi}),  \\
\end{equation}
where $r_0$ is the equilibrium position of  each carbon atom, and $A$
is the  amplitude. In the calculation, we take $\varphi=\pi/2$ and
$A$ to be 5 percent of the intersection radius. This percentage is
significantly larger than the one in the
literature~\cite{long05,rara10,so00,ku03}, and we choose it for the
following considerations. Though this work focuses on the RBM effect,
one may be more interested in the energy transfer or molecule
transport process of the microsystems induced by the more general and
complex collective vibrations that involve the RBM and usually have
much larger amplitudes~\cite{rao97,kahm99,gibs07}. On the other hand,
the vibration amplitude of a biological microtubule can also be much
larger~\cite{wa08,poko04}.  Specifically, we have tested that the
(6,6) SWCNT is rather flexible and it is not very difficult to reach
the vibration amplitude as much as 5 \% by appropriate external
induction~\cite{alex07,ronald07}. Without losing the generality, we
thus choose a moderate amplitude 5\% to observe clear effects of
collective vibrations and provide useful insights for the simulations
of the biological microsystems. Moreover, because the filling of the
water just affects the RBM frequency slightly~\cite{camb10}, we
neglect it in our estimation.

\emph{Results and Discussion}--- The simulation is performed for the
SWCNT system with different RBM frequencies.  For each frequency, the
simulation time is 110 ns. Unless otherwise indicated, the data of the
last 100 ns is used for analysis. For clarity, we define the net flux
to be the difference between the water molecular number per
nanosecond across the SWCNT from one end to the
other~\cite{gon08,wan05}. The average number~\cite{hu01,gon08} of
water molecules inside SWCNT is denoted with the symbol $\bar{N}$.

\begin{figure}[thb]
\vspace*{-5mm}
\includegraphics[height=8cm,width=9.3cm]{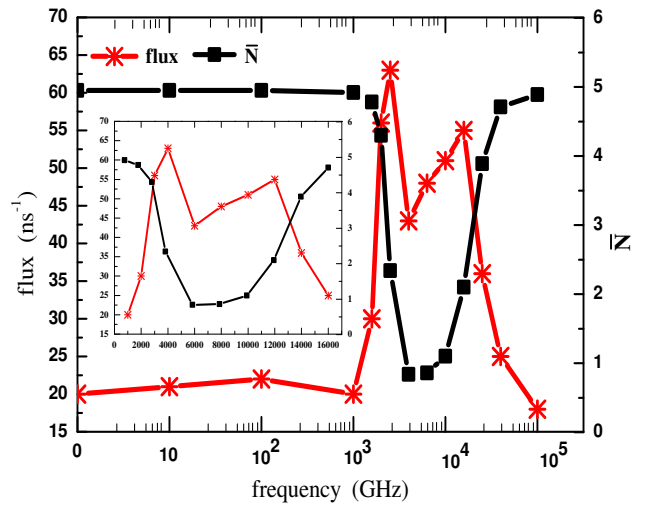}
\vspace*{-10mm} \caption{(Color online) The net flux and average
number $\bar{N}$ of water molecules inside the tube as a function of
RBM frequency. Given in the inset is an image amplification within
the frequency interval 0-16000 GHz. \label{fflux}}
\end{figure}

Fig.~\ref{fflux} shows  the net flux and $\bar{N}$ of water molecules
across the SWCNT as a function of RBM frequency. It is observed that
the net flux and $\bar{N}$ do not change much at most frequencies and
are about 20 ns$^{-1}$ and 5, respectively. However, as shown in
Fig.~\ref{fflux} (inset), the net fluxes undergo a clear enhancement
in the frequency range from about  2000 GHz to 14000 GHz, and two
peak values around 4000 GHz and 12000 GHz are 63 ns$^{-1}$ and 55
ns$^{-1}$, respectively, being approximately three times those beyond
this frequency range.   As shown in Fig.~\ref{fflux}, we observe that
the $\bar{N}$ is around 1 in the frequency range of 5000-10000 GHz,
and its minimum is about 0.84 at 6000 GHz. Water molecules inside the
SWCNT are actually absent within 5000-10000 GHz in a majority of
time. As an example, we have counted up the temporal distribution of
the average number of water molecules inside the SWCNT in the 100 ns
at 10$^{4}$ GHz. We find that the empty state, namely, no water
molecule inside the tube, accounts for 46.1\% of total time, and the
occupancy with $\bar{N}=1,2,3,4,$ and 5 accounts for 22.5\%, 14.5\%,
9.9\%, 5.5\%, and 1.5\%, respectively. Meanwhile, we have counted the
hydrogen bond number of water molecules inside the SWCNT in this
frequency range, according to the criterion of the oxygen distance
less than 3.5 $\overset{\circ}{A}$ and hydrogen-bond angle $\leq
30^{\circ}$~\cite{wan05}, and consistently found that it is almost
always zero. Clearly, with hydrogen bonds of water molecules being
almost completely broken, the water molecules move individually in
the SWCNT with the RBM in this frequency range, in contrast to the
existence of the single-file water chain at other frequencies.
Consequently, the sharp enhancement of the flux follows from the full
fracture of hydrogen bonds and fast shuttling of water molecules from
one end to the other.

A coincident breaking of the hydrogen bonds can be realized by a
perturbation of the form $e^{i\omega t}$ in sufficiently long time
span, while the understanding of a thorough fracture can resort to a
classical harmonic oscillator model. Supposing the intrinsic
frequency of the single-file water chain is determined by the binding
energy of the hydrogen bond, $E_h$, we can estimate the classical
resonant frequency using the relation $E_h=m\omega^2A^2/2$ where $m$
is the mass of water molecule, and $A$ is the amplitude of allowed
radial motion. Using the value of $E_h=16 $ kcal/mol~\cite{hu01} and
the amplitude about 0.3-0.9 $\overset{\circ}{A}$ which are half-peak
width values of the Gaussian distribution observed in the simulation,
the classical resonant frequency of water chain is estimated to lie
between 4800 GHz and 14500 GHz which is well situated in the
frequency range for large fluxes in our simulations. It is noteworthy
that the theoretical value of inherent frequency of the short (6,6)
SWCNT is about $10^4$ GHz ~\cite{law05,long05,chen10,geor09}. With
the RBM frequency away from the resonance, segments of the water
chain instead of individual molecules are produced and hence the
occupancy of water molecules increases. A double-peak structure of
the water flux thus appears due to the fast shuttling of the broken
segments. With the RBM frequency far away from the resonance, the
flux is almost unaffected by the RBM because there is just coincident
breaking of hydrogen bonds. Experimentally, this drastic resonant
effects on the water permeation can be verified with radio wave
techniques.

\begin{figure}[thb]
\begin{flushleft}
\vspace*{-10mm}
\includegraphics[height=9cm,width=9.0cm]{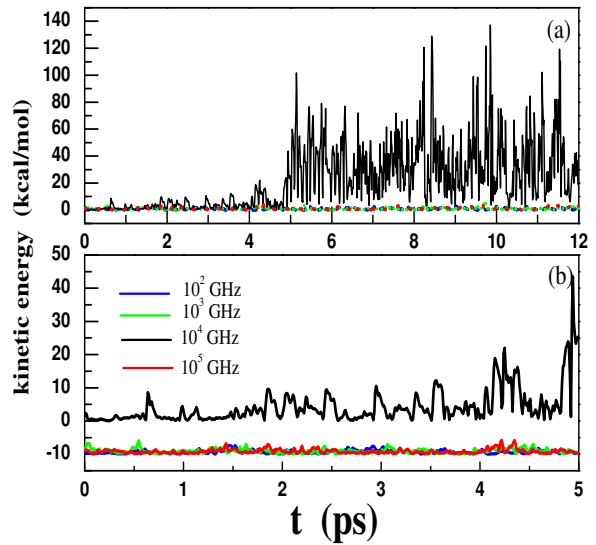}
 \end{flushleft}
\vspace*{-10mm} \caption{(Color online) The kinetic energy for each
water molecule inside the SWCNT with respect to time. Results in an
earlier period are presented in the lower panel where the colored
curves are displaced downwards for clarity. \label{fke}}
\end{figure}

To better understand  the process of the hydrogen bond fracture, a
dynamic evolution of the average kinetic energy of water molecules
held in the tube is shown in Fig.~\ref{fke}. It is interesting to see
that the significant fluctuation at $f=10^4$ GHz starts to form
actually at a very early time  and then the shift to a strong
oscillation occurs after a relatively long latent period. While this
phenomenon does not take place at other frequencies as shown in
Fig.~\ref{fke}, it can be understood as a resonant response to the
RBM of the SWCNT. We note that this resonance accompanied with
the H-bond rupture needs an energy accumulation process with the time
span depending on the RBM amplitude. 

\begin{figure}[thb]
\begin{center}
\vspace*{-5mm}
\includegraphics[height=8cm,width=9.0cm]{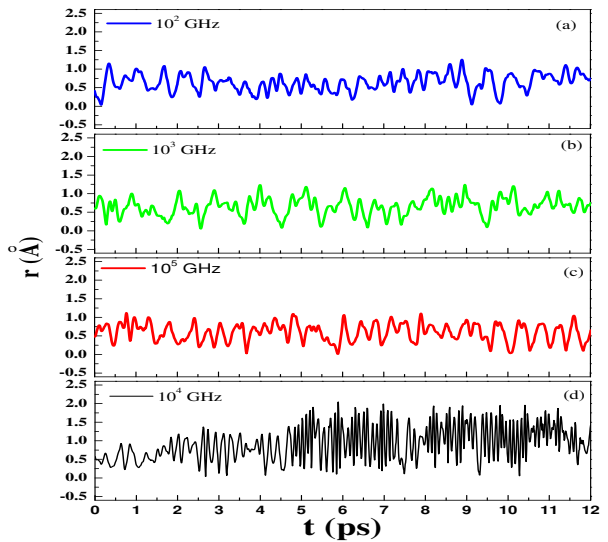}
 \end{center}
\vspace*{-5mm} \caption{(Color online) The distance of the oxygen
atom of the water molecule from the z-axis evolving with time for
various RBM frequencies. \label{fdis}}
\end{figure}

\begin{figure}[thb]
\vspace*{-5mm}
\includegraphics[height=9cm,width=9.0cm]{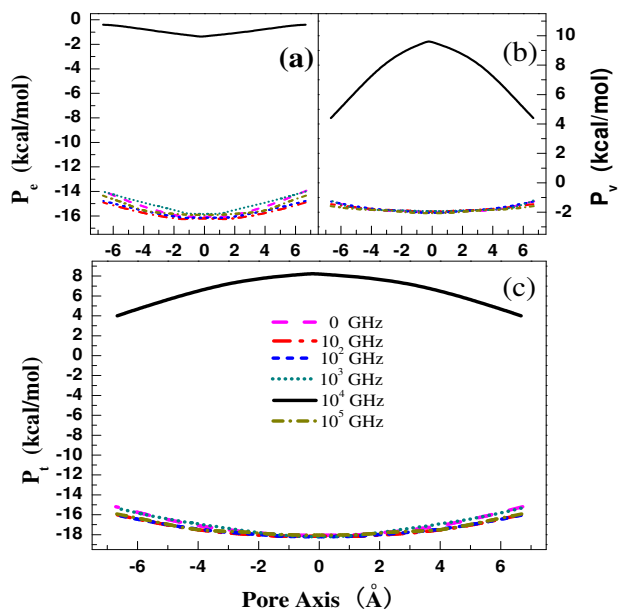}
\vspace*{-8mm} \caption{(Color online) The average electrostatic (a),
van der Waals (b), and total (c) potentials per water molecule that
interacts with the SWCNT, two graphite sheets, and all ambient water
molecules as a function of z-coordinates. \label{finp}}
\end{figure}

Shown in Fig.~\ref{fdis} is the distance of the oxygen atom of the
water molecule inside the SWCNT from the tube axis evolving with the
time. Similarly, we see again that the oscillation is forced to
resonate with the RBM of the SWCNT at 10$^{4}$ GHz, while the
coherent oscillations do not occur at other frequencies. Posterior to
the fracture of hydrogen bonds, the radial oscillation is dictated by
the van der Waals interaction between the carbon atoms of the SWCNT
and water molecules.
It is known that the motion of water molecules is driven by the
potential field. We thus calculate the average electrostatic, van der
Waals and total interaction potentials: $P_{e}$, $P_{v}$ and
$P_t=P_e+P_v$ per water molecule at different position along z axis
inside the tube where the water molecule interacts with the SWCNT,
two graphite sheets, and all ambient water molecules. As shown in
Fig.~\ref{finp}, the average interaction potentials are almost the
same at all frequencies but $f=10^4$ GHz. This explains why the net
flux and $\bar{N}$ do not have significant differences at most
frequencies. In fact, we can continuously observe stable
one-dimensionally ordered water chain like previous
works~\cite{hu01,wan05,gon08} using molecular visualization program
VMD~\cite{hum96} at these frequencies. However, at 10$^{4}$ GHz,
Fig.~\ref{finp} (a) shows that the $P_{e}$ is close to 0 kcal/mol,
which means that hydrogen bonds of water molecules inside the SWCNT
are almost completely deprived. Fig.~\ref{finp} (b) shows that the
$P_{v}$ is repulsive for single water molecules, revealing the fact
that the single water molecules are forced to move closely to the
tube wall. The $P_{t}$, as shown in Fig.~\ref{finp}(c), demonstrates
a similar character featured by $P_e$ and $P_v$. Due to the repulsion
provided by the RBM resonance, it is difficult for water molecules to
enter into and keep in the SWCNT, resulting in a dramatic reduction
of $\bar{N}$, whereas this can indeed provide an efficient transport
of water molecules accompanied by a large net flux. We mention that
the peak values of the net flux do not coincide  with the minimum
values of $\bar{N}$ as discussed above. On the other hand, we note
that the water molecules can  be as far as 2 Angstrom from the z axis
when the resonance happens (see Fig.~\ref{fdis}). In such cases, it
is necessary to explore the validity of the force-field model. With
comparison to the total energy of the snapshots obtained from the
ab-initio approach~\cite{li12}, we find that the energy acquirement
of water molecules is clearly overestimated by the MD simulation. By
reducing the radius parameters in the MD simulation, we may roughly
reproduce the ab-initio results. In doing so, the phenomena of
resonance remain unchanged, while the flux is just moderately
enhanced. We observe that the H-bond fracture due to the resonant
mechanism are mainly relevant to the relatively long-range van der
waals interaction in the vicinity of the equilibrium position, while
the short-range repulsive interaction plays a more efficient role
posterior to the H-bond fracture. Therefore, the main conclusions
that we have drawn are not essentially affected by the deficiency of
the MD method.

It is worth discussing the dependence of the flux on the RBM
amplitude. The RBM effect on the flux  decreases with the decrease of
the RBM amplitude. For instance, for an amplitude of 3\% of the
intersection radius,  the flux at $f=10^4$ GHz reduces to be about 40
ns$^{-1}$ which is just twice the value at frequencies without  the
RBM influence. The flux enhancement can further be reduced till to
disappearance at a less RBM amplitude. With the present setup, the
smallest RBM amplitude is about 2.7\%, while it can reduce moderately
with decreasing the imposed pressure gradient.

At last, it is significant to explore the RBM effects on the water
transport in SWCNTs with generalized setups that may bridge the
connection with the experiments using the larger diameters and longer
lengths. We have chosen the SWCNT setups with the diameter ranging
from 1.09 to 1.36 nm and the length from 1.34 to 10 nm. Likewise, we
have observed that the  rupture of the H-bonds of various
hinged water chains can occur in all these SWCNTs with the RBM at
resonant frequencies, resulting in a noticeable difference in water
fluxes. Moreover, we  have studied the RBM effects on water transport
properties at  much smaller pressure gradients. It is interesting to
observe that the water chain fracture is more remarkable for a small pressure gradient, e.g., 15 atm,
owing to a much longer stay of water molecules in the tube. These
investigations indicate that the RBM resonance is rather universal.

\emph{Conclusions}---In summary, we have demonstrated  effects of the
resonant mechanism, responding to the RBM of the SWCNT, on the
transport properties and dynamical behaviors of water molecules
across the SWCNT. The resonance of the single-file water chain with
the RBM of the SWCNT, characterized by the full fracture of hydrogen
bonds, results in a dramatic decrease of water molecules inside the
tube and a sharp enhancement of the net flux. It is also found that
away from the resonance on the both sides roughly in a frequency
range 2000-14000 GHz the RBM of the SWCNT can create partial breaks
of hydrogen bonds and accordingly dramatic changes in transport
properties of water molecules, whereas the latter are almost
unaffected beyond this range.  In this frequency range, a double-peak
structure of the net flux forms with the vast enhancement of the net
flux, and the peak  values around 4000 GHz and 12000 GHz are three
times those unaffected by the RBM. Our findings may have some
possible implications to biological activities and potential
technical applications in the development of nanotube-based
nanofluidic devices.

The work was supported in part by the National Natural Science
Foundation of China under Grant Nos. 10975033 and 11275048, the China
Jiangsu Provincial Natural Science Foundation  under Grant
No.BK2009261, and Natural Science Foundation of Anhui Province of
China under Grant No.1208085QA16.


\begin{thebibliography}{99}

\bibitem{mur00}K. Murata, K. Mitsuoka, T. Hirai, T. Walz, P. Agre, J. B. Hemann, A.
Engel, and Y. Fujiyoshi, Nature (London) \textbf{407}, 599 (2000).
\bibitem{zhu04}F. Q. Zhu, E. Tajkhorshid, and K. Schulten, Phys. Rev. Lett. \textbf{93}, 224501 (2004).
\bibitem{gon08}X. J. Gong, J. Y. Li, H. J. Lu, H. Zhang, R. Z. Wan, and H. P. Fang, Phys. Rev. Lett. \textbf{101}, 257801 (2008);
X. J. Gong, J. C. Li, K. Xu, J. F. Wang, and H. Yang, J. Am. Chem.Soc. \textbf{132}, 1873 (2010).
\bibitem{son08} S. Joseph and N. R. Aluru, Nano Lett. \textbf{8}, 452 (2008).
\bibitem{liu10}J. Liu, J. F. Fan, M. Tang, Min Cen, J. F. Yan, Z. Liu, and W. Q. Zhou, J. Phys. Chem. B \textbf{114}, 12183 (2010).
\bibitem{duan10}W. H. Duan and Q. Wang, ACS Nano. \textbf{4}, 2338 (2010).
\bibitem{hol06}J. K. Holt \emph{et al.}, Science \textbf{312}, 1034 (2006).
\bibitem{lon07}M. J. Longhurst and  N. Quirke, Nano Lett. \textbf{7}, 3324 (2007).
\bibitem{li07}J. Y. Li, X. J. Gong, H. J. Lu, D. Li, H. P. Fang, and R. H. Zhou, Proc. Natl. Acad. Sci. U.S.A. \textbf{104}, 3687 (2007).
\bibitem{wang06}B. Wang and P. Kra, J. Am. Chem. Soc. \textbf{128} 15984 (2006).
\bibitem{jia11}J. Y. Su and H. X. Guo, ACS Nano. \textbf{5}, 351 (2011).
\bibitem{corr08}B. Corry, J. Phys. Chem. B \textbf{112}, 1427 (2008).
\bibitem{jira97}K. B. Jirage, J. C. Hulteen, and C. R. Martin, Science \textbf{278}, 655 (1997).
\bibitem{gon07}X. J. Gong, J. Y. Li, H. J. Lu, R. Z. Wan, J. C. Li, J. Hu, and H. P. Fang, Nat. Nanotechnol. \textbf{2}, 709 (2007).
\bibitem{groo01}B. L. de Groot and H. Grubmu¡§ller, Science \textbf{294}, 2353 (2001).
\bibitem{hu01}G. Hummer, J. C. Rasaiah, and J. P. Noworyta, Nature (London) \textbf{414}, 188 (2001).
\bibitem{anis04}A. Anishkin and S. Sukharev, Biophys. J. \textbf{86}, 2883 (2004).
\bibitem{beck03}O. Beckstein and M. S. P. Sansom, Proc. Natl. Acad. Sci. U.S.A.  \textbf{100}, 7063 (2003).
\bibitem{wan05}R. Z. Wan, J. Y. Li, H. J. Lu, and H. P. Fang, J. Am. Chem.Soc. \textbf{127}, 7166 (2005).
\bibitem{lot1}O. Beckstein, \emph{et al.}, J. Phys.Chem. B \textbf{105}, 12902 (2001); A. V. Raghunathan
and N. R. Aluru, Phys. Rev. Lett. \textbf{97},024501 (2006); P. Gallo, \emph{et al.}, J. Chem. Phys. \textbf{113}, 11324 (2000); L. Kullman, \emph{et al.}, Biophys. J. \textbf{82}, 803 (2002).
\bibitem{rao97}A. M. Rao, \emph{et al.}, Science, \textbf{275},188 (1997).
\bibitem{dub03}O. Dubay and G. Kresse, Phys. Rev. B. \textbf{67}, 035401 (2003).
\bibitem{law05}H. M. Lawler, D. Areshkin, J. W. Mintmire, and C. T. White,
Phys. Rev. B. \textbf{72}, 233403 (2005).
\bibitem{yan07}W. Yang, R. Z. Wang, X. M. Song, B. Wang, and H. Yan, Phys. Rev. B. \textbf{75}, 045425 (2007).
\bibitem{arau10}P. T. Araujo, P. B. C. Pesce, M. S. Dresselhaus, K. Sato, R. Saito, and A. Jorio,  Physica E. \textbf{42} 1251 (2010).
\bibitem{zuo10}G. C. Zuo, R. Shen, S. J. Ma, and W. L. Guo, ACS Nano. \textbf{4}, 205 (2010).
\bibitem{dar93} T. A. Darden, D. M. York, and L. G. Pedersen, J. Chem. Phys. \textbf{98}, 10089 (1993).
\bibitem{jam05} James C.  Phillips \emph{et al.}, J. Comput. Chem. \textbf{26}, 1781 (2005).
\bibitem{mac98}A. D. MacKerell \emph{et al.}, J Phys Chem B \textbf{102}, 3586 (1998).
\bibitem{zhu02} F. Q. Zhu, E. Tajkhorshid, and K. Schulten, Biophys. J. \textbf{83}, 154 (2002).
\bibitem{jor83}W. L. Jorgensen, J. Chandrasekhar, J. D. Madura, R. W. Impey, and M. L. Klein, J. Chem. Phys. \textbf{79}, 926 (1983).
\bibitem{long05}M. J. Longhurst and N. QUIRKE, Molecular Simulation \textbf{31}, 135 (2005).
\bibitem{rara10} N. R. Raravikar P. Keblinski, A. M. Rao, M. S. Dresselhaus, L. S. Schadler, and P. M. Ajayan, Phys. Rev. B. \textbf{66}, 235424 (2002).
\bibitem{so00} V. P. Sokhan, D. Nicholson, and N. Quirke, J. Chem. Phys. \textbf{113}, 2007 (2000).
\bibitem{ku03}J. K\"uti, V. Zolyomi, M. Kertesz, and G. Sun, New J. Phys. \textbf{5}, 125.1 (2003).

\bibitem{kahm99}D. Kahn and J. P. Lu, Phys. Rev. B \textbf{60}, 6535 (1999).
\bibitem{gibs07}R. F. Gibson, E. O. Ayorinde, and Y. F. Wen, Compos. Sci. and Tech., \textbf{67} 1 (2007).
\bibitem{alex07}P. Alex Greaney and Jeffrey C. Grossman, Phys. Rev. Lett \textbf{98}, 125503 (2007).
\bibitem{ronald07}Ronald F. Gibson a,Emmanuel O. Ayorinde a, and Yuan-Feng Wen, Composites Science and Technology \textbf{67} 1 (2007).

\bibitem{wa08}C. Y. Wang and L. C. Zhang, J. biomech. \textbf{41}, 1892 (2008).
\bibitem{poko04}J. Pokorn\'y, Bioelectrochem. \textbf{63}, 321 (2004).
\bibitem{camb10}S. Cambr\'e \emph{et al.}, Phys. Rev. Lett. \textbf{104}, 207401 (2010).
\bibitem{chen10}H. C. Cheng \emph{et al.}, Comput. Methods Appl. Mech. Engrg. \textbf{199}, 2820 (2010).
\bibitem{geor09}S. K.  Georgantzinos, G. I. Giannopoulos and N. K. Anifantis, Comput Mech \textbf{43}, 731 (2009).
\bibitem{hum96}W. Humphrey, A. Dalke, and K. Schulten, J. Molec. Graphics \textbf{14.1}, 33 (1996).
\bibitem{li12}X. Li \emph{et al.},Phys. Rev. B. \textbf{85}, 085425 (2012).

\end{thebibliography}
\end{document}